\documentclass[twocolumn,10pt,superscriptaddress,nofootinbib,aps,prl]{revtex4} % reprint

\usepackage[utf8]{inputenc}

\usepackage{hyperref}
\usepackage{braket}
\usepackage{upgreek}
\usepackage[normalem]{ulem}
\usepackage[dvips]{graphicx}
\usepackage{hhline}
\usepackage{epsfig,amsmath,amssymb,verbatim,mathrsfs,array,layout,textcomp,latexsym,slashed,graphicx,booktabs,color,mathtools}
\usepackage[caption=false]{subfig} % added to put figures next to each other. The caption=false flag avoids conflicts with revtex. - AI

\makeatletter
\DeclareFontFamily{OMX}{MnSymbolE}{}
\DeclareSymbolFont{MnLargeSymbols}{OMX}{MnSymbolE}{m}{n}
\SetSymbolFont{MnLargeSymbols}{bold}{OMX}{MnSymbolE}{b}{n}
\DeclareFontShape{OMX}{MnSymbolE}{m}{n}{
    <-6>  MnSymbolE5
   <6-7>  MnSymbolE6
   <7-8>  MnSymbolE7
   <8-9>  MnSymbolE8
   <9-10> MnSymbolE9
  <10-12> MnSymbolE10
  <12->   MnSymbolE12
}{}
\DeclareFontShape{OMX}{MnSymbolE}{b}{n}{
    <-6>  MnSymbolE-Bold5
   <6-7>  MnSymbolE-Bold6
   <7-8>  MnSymbolE-Bold7
   <8-9>  MnSymbolE-Bold8
   <9-10> MnSymbolE-Bold9
  <10-12> MnSymbolE-Bold10
  <12->   MnSymbolE-Bold12
}{}

\let\llangle\@undefined
\let\rrangle\@undefined
\DeclareMathDelimiter{\llangle}{\mathopen}%
                     {MnLargeSymbols}{'164}{MnLargeSymbols}{'164}
\DeclareMathDelimiter{\rrangle}{\mathclose}%
                     {MnLargeSymbols}{'171}{MnLargeSymbols}{'171}
\makeatother

\usepackage{microtype} % for bad boxes
\newcommand{\beq}{\begin{equation}}
\newcommand{\eeq}{\end{equation}}

\def\beqa{\begin{eqnarray}}
\def\eeqa{\end{eqnarray}}
\def\bea{\begin{eqnarray}}
\def\eea{\end{eqnarray}}
\newcommand{\be}{\begin{eqnarray}}
\newcommand{\ee}{\end{eqnarray}}
\newcommand{\bv}{\left(\begin{array}{c}}
\newcommand{\ev}{\end{array}\right)}
\newcommand{\bmtwo}{\left(\begin{array}{cc}}
\newcommand{\bmthree}{\left(\begin{array}{ccc}}
\newcommand{\emn}{\end{array}\right)}
\newcommand{\bmtwoc}{\left\{\begin{array}{cc}}
\newcommand{\bmthreec}{\left\{\begin{array}{ccc}}
\newcommand{\emnc}{\end{array}\right\}}
\newcommand{\ba}{\begin{array}}
\newcommand{\ea}{\end{array}}

\definecolor{readableRTD}{rgb}{0.7,0.1,0.2}

\newcommand{\bk}[1]{\left<#1\right>}
\newcommand{\bks}[1]{\left[#1\right]}

\newcommand{\kett}[1]{\left|#1\right\rrangle}
\newcommand{\braa}[1]{\left\llangle #1\right|}
\newcommand{\sket}[1]{\left|#1\right]}
\newcommand{\sbra}[1]{\left[#1\right|}
\newcommand{\skett}[1]{\left|\left.#1\right]\right]}
\newcommand{\sbraa}[1]{\left[\left[#1\right.\right|}

\newcommand{\colmatt}[1]{\left(\begin{array}{cc}#1\end{array}\right)}

\def\lsim{\mathrel{\rlap{\lower4pt\hbox{\hskip1pt$\sim$}}
     \raise1pt\hbox{$<$}}}         % less than or approx. symbol
\def\gsim{\mathrel{\rlap{\lower4pt\hbox{\hskip1pt$\sim$}}
     \raise1pt\hbox{$>$}}}         % greater than or approx. symbol

\begin{document}

%\vspace*{-30mm}
%\font\mini=cmr10 at 0.8pt

\title{
Partially Celestial States and Their Scattering Amplitudes
}

\author{Csaba Cs\'aki}
\affiliation{Department of Physics, LEPP, Cornell University, Ithaca, NY 14853, USA}
\author{Ofri Telem}
\affiliation{Racah Institute of Physics, Hebrew University of Jerusalem, Jerusalem 91904, Israel}
\author{John Terning}
\affiliation{QMAP, Department of Physics, University of California, Davis, CA 95616, USA}

\begin{abstract}We study representations of the Poincar\'e group that have a privileged transformation law along a $p$-dimensional hyperplane, and uncover their associated spinor helicity variables in $D$ spacetime dimensions. Our novel representations generalize the recently introduced celestial states and transform as conformal primaries of $SO(p,1)$, the symmetry group of the $p$-hyperplane. We will refer to our generalized states as ``partially celestial." Following Wigner's method, we find the induced representations, including spin degrees of freedom. Defining generalized spinor helicity variables for every $D$ and $p$, we are able to construct the little group covariant part of partially celestial amplitudes. Finally, we briefly examine the application of the pairwise little group to partially celestial states with mutually non-local charges.
\end{abstract}

\maketitle
\section{Introduction}

The classification of particles into representations of the Poincar\'e group is the basis of particle physics and quantum field theory (QFT): it allows for the definition of the $S$-matrix and scattering amplitudes and the easy identification of the propagating degrees of freedom of QFT's. The little group appearing in Wigner's method of induced representations is essential for building proper scattering amplitudes and forms the basis of modern scattering amplitude methods. While little group methods have by now become commonly used for particle scattering, it has not been widely applied to the description of the dynamics of branes. In this paper we initiate the first steps toward this direction. 
%Over the past three decades $p$-branes have played an ever more prominent role in high energy physics and string theory. In their guise as D-branes, they provided the foundation of string dualities, the AdS/CFT correspondence, modern black hole physics and RS/holographic model building, just to name a few examples. From the point of view of supergravity (SUGRA), $p$-branes are solitonic objects, and it should be possible to quantize them similarly to other solitons in quantum field theory. 
While we do not consider the seemingly formidable task of quantizing $p$-branes, we will consider a simpler situation where brane-like objects appear and their general states can be constructed using Wigner's method. To achieve this we will define a new eigenbasis of ordinary quantum fields in $D$ dimensions that have privileged transformation properties on a $p$-hyperplane. We call this state a $p$\textit{-sheet} or a $p$\textit{-partially-celestial} state, for reasons that will become obvious below. Though not quite a $p$-brane, the $p$-sheet does serve as an interesting toy-model for $p$-branes, as it highlights the importance of $SO(D-p-1)$ transverse rotations, a feature that we expect to play a key role in a future ``Wigner" quantization of $p$-branes. 

Our starting point will be to look for states that (a) have well-defined $SO(p,1)$ transformation properties, reflecting the symmetry of a $p+1$-worldsheet; and (b) are not \textit{zero-energy eigenstates}. In fact, these two requirement imply that our $p$-sheet states are not energy eigenstates at all. As we shall see in detail below, the $SO(p,1)$ covariance of our $p$-sheets makes them the analogs of the celestial states considered in~\cite{Pasterski2017a,Pasterski2017b,,Pasterski2017,Kapec2018,Banerjee2020,Banerjee2020a,Pasterski2021}, except only along $p$ directions, hence they are ``partially celestial''. We will find the appropriate eigenbasis of these states, and also find the correct labels for characterizing $p$-sheet quantum states.  With our knowledge of the little group and canonical Lorentz transformations we can use Wigner's method of induced representations to build up the full $p$-sheet Hilbert space. We are also able to present for the first time  the generalized spinor helicity variables in any spacetime dimension, which has applications far beyond those presented here, and is the most far reaching result in this paper. These variables allow us to construct the most general 3-point amplitudes for partially celestial states. We also briefly consider  how the recently introduced pairwise little group \cite{Csaki:2020inw,Csaki:2020yei,Csaki2022a,Csaki2022} can be generalized to $p$-sheets. For the case of mutually non-local sheets of dimension $p$ and $D-p-4$ we show that the pairwise little group is just a $U(1)$, providing a new example of pairwise helicity, which can be dynamically realized if $p+1$-form electrodynamics is electrically coupled to the $p$-sheet and magnetically coupled to the dual $D-p-4$ sheet. 

The paper is organized as follows. First we briefly review the celestial solutions \cite{Pasterski2017a,Pasterski2017b,,Pasterski2017,Kapec2018,Banerjee2020,Banerjee2020a,Pasterski2021} of the Klein-Gordon (KG) equation, which are solutions that transform covariantly with respect to $SO(D-1,1)$, viewed as the Euclidean conformal group. Using celestial solutions as an inspiration, we then present solutions of the $d$-dimensional KG equation which are $SO(p,1)$ covariant, reflecting the symmetry of a $p+1$-worldvolume. In their ``rest" frame, these solutions are also manifestly $SO(D-p-1)$ rotationally invariant and $R^{D-p-1}$ translationally invariant in the space orthogonal to the $p$-sheet. Next, we show how to interpret these $SO(p,1)$ covariant solutions of the KG equation as the \textit{wavefunctions} of $p$-partially-celestial quantum states, thus constructing their Hilbert space. The generalization to spinning $p$-sheets is then achieved using Wigner's method of induced representations.
We then construct spinor helicity variables that allow us to write the most general 3-point amplitudes. Finally we present the pairwise little group of two parallel sheets, and argue that for mutually non-local sheets the pairwise little group reduces to a $U(1)$ pairwise helicity. 

\section{Plane Waves and Celestial Scalars}\label{sec:PWCel}

In preparation for presenting our $p$-sheet states we will first review the construction of the celestial scalars and their relation to plane waves. Consider first a massive classical scalar field $\phi(x)$ in $D$-dimensions. Its equation of motion is the Klein-Gordon (KG) equation (in a mostly-plus signature as is commonly used in the celestial literature) 
\begin{eqnarray}\label{eq:weq}
\left[-\partial^2_t+\nabla^2+m^2\right]\phi(x)=0\,.
\end{eqnarray}
The most commonly used basis of solutions is the plane wave basis $\phi_{p}(x)=e^{\pm ip\cdot x}$. Each solution $\phi_{p}(x)$ in this basis is translationally invariant in $D-1$ directions $x^\mu\rightarrow x^\mu+ \Delta {x}^\mu$ orthogonal to $p^\mu$, $p\cdot\Delta x=0$. One could instead look for solutions of \eqref{eq:weq} which are $SO(D-1,1)$ covariant --- these are the massive celestial scalars \cite{Pasterski2017a,Law2020} $\phi_\Delta(x;\vec{w})$. Instead of the $p^\mu$ labels, these solutions are labeled by a conformal dimension $\Delta$ and a vector $\vec{w}$ on $R^{d}$, where $d\equiv D-2$. Explicitly, they are given by
\begin{eqnarray}\label{eq:solsod1}
&&\phi^{\pm,\,PS}_{\Delta}(x;\vec{w})=\frac{2^{\frac{d}{2}+1}\pi^{\frac{d}{2}}}{(im)^{\frac{d}{2}}}\,\frac{(is)^{\alpha}}{\left[-q(\vec{w})\cdot x\mp i\epsilon)\right]^\Delta}\,K_{\alpha}(ms)\nonumber\\[5pt]
&&s=\sqrt{x\cdot x},~~~\alpha=\Delta-\frac{d}{2}\,.
\end{eqnarray}
The label PS here is to remind us that these are the celestial wavefunctions defined in \cite{Pasterski2017a}. Here
\begin{eqnarray}\label{eq:qdef}
q^\mu(\vec{w})=(1+|\vec{w}|^2,2\vec{w},1-|\vec{w}|^2),
\end{eqnarray}
is a $D=d+2$ dimensional vector. A Lorentz transformation $\Lambda$ acting on $q^\mu$ induces a nonlinear map $\Lambda:\vec{w}\rightarrow\vec{w}'$ via
\begin{eqnarray}\label{eq:qdef}
q^\mu(\vec{w}')={\left|\frac{\partial\vec{w}'}{\partial\vec{w}}\right|}^{1/d}\Lambda^\mu_{~\nu}\,q^\nu(\vec{w})\,.
\end{eqnarray}
The map $\Lambda:\vec{w}\rightarrow\vec{w}'$ non-linearly realizes $SO(d-1,1)$ as the Euclidean conformal group acting on $\vec{w}\in R^{d-2}$. By substituting \eqref{eq:qdef} in \eqref{eq:solsod1}, one can easily check that these solutions have the property that
\begin{eqnarray}\label{eq:trcov}
\phi_\Delta\left(\Lambda^\mu_\nu x^\nu,\vec{w}'(\vec{w})\right)={\left|\frac{\partial\vec{w}'}{\partial\vec{w}}\right|}^{-\frac{\Delta}{d}}\phi_{\Delta}\left(x^\nu,\vec{w}\right)\,,
\end{eqnarray}
The solutions \ref{eq:solsod1} form a complete eigenbasis for the KG equation for either $\Delta=\frac{d}{2}+i\mathbb{R}$ or $0<\Delta<1$, also called the \textit{principal series} and \textit{complementary series} representations of $SO(D-1,1)$, respectively.  

The celestial wavefunctions for massless scalars were obtained in \cite{Pasterski2017a} by taking the massless limit of \eqref{eq:solsod1}. In \cite{Banerjee2019}, an equivalent construction of the celestial states for massless particles in $4D$ was presented. The latter followed Wigner's method of induced representations, by starting from a reference quantum state whose little group is the ``lower triangular" group $\left\{J_3,K_3,J_2-K_1,-J_1-K_2\right\}$. Inspired by this construction, we present a slightly modified derivation of the solution \eqref{eq:solsod1} using little group methods (see also a parallel discussion for the massless case in the very recent \cite{Donnay2022}), which will be easily generalized to our partially celestial $p$-sheet solutions. First, we redefine the expression in \eqref{eq:solsod1}, as a function of $q^\mu$ rather than $\vec{w}$,
\begin{eqnarray}\label{eq:solsod2}
&&\phi^{\pm}_{\Delta}(x;q)=\frac{2^{\frac{d}{2}+1}\pi^{\frac{d}{2}}}{(im)^{\frac{d}{2}}}\,\frac{(is)^{\alpha}}{\left[-q\cdot x\mp i\epsilon)\right]^\Delta}K_{\alpha}(ms)\nonumber\\[5pt]
&&s=\sqrt{x\cdot x},~~~\alpha=\Delta-\frac{d}{2}\,.
\end{eqnarray}
Note that this is a solution to the KG equation \textit{only for null} $q^\mu$. Naturally, we define a reference value of $q^\mu$ as
\begin{eqnarray}\label{eq:qref}
&&q^\mu_{{\rm ref}}\equiv \left(1,0,\ldots,0,1\right)\,,
\end{eqnarray}
and note that this is a lightlike Lorentz $D$-vector as opposed to the $D-2$ vector $\vec{w}$.  Accordingly, a ``reference wavefunction" is
\begin{eqnarray}\label{eq:phiref0}
&&\phi^{\pm,{\rm ref}}_\Delta(x)\equiv\phi^{\pm}_\Delta(x;q_{{\rm ref}})\,.
\end{eqnarray}
Here $q^\mu_{{\rm ref}}$ is chosen so that $\phi^{\pm,{\rm ref}}_\Delta(x)$ is $SO(D-2)$ rotationally invariant, but it is also manifestly invariant under $M_{i,D-1}-M_{0,i},\,i\in\{1,\ldots,D-1\}$. Overall, the little group under which $\phi^{\pm,{\rm ref}}_\Delta(x)$ is invariant is given by
\begin{eqnarray}\label{eq:LGcel}
LG^D=ISO(D-2)\,,
\end{eqnarray}
where $dim(LG^D)=\frac{(D-1)(D-2)}{2}$. Under general transformations in the Poincar\'e  group $\mathcal{P}^D=\mathbb{R}^D\rtimes SO(D-1,1)$, the celestial scalar transforms as 
\begin{eqnarray}\label{eq:psscalcel}
&&\Omega=(\Lambda,v)\in\mathcal{P}^D:\nonumber\\[5pt]
&&\phi^{\pm, {\rm ref}}_\Delta(x)\rightarrow \phi^{\pm, \,\Omega}_\Delta(x)=\phi^{\pm, {\rm ref}}_\Delta(\Lambda(x+v))\nonumber\\[5pt]
&&=\phi^{\pm}_\Delta(x+v;\Lambda^{-1} q)\,.
\end{eqnarray}
Without loss of generality, we perform the translation before the Lorentz transformation. Note also that we could further Taylor-expand the last line of \eqref{eq:psscalcel} in $v^\mu$, and see that the action of an internal translation $P^\mu$ shifts $\Delta\rightarrow\Delta+1$ \cite{Donnay2019,Stieberger2019,Law2020a}, but we will not do it explicitly in this paper.
Importantly, not all Poincar\'e transformations actually lead to inequivalent solutions for $\phi$ --- to label inequivalent solutions we have to mod out by the action of the little group. This is the same as finding canonical transformations $O\in \mathcal{P}^D/LG^D$. 
From \eqref{eq:psscalcel}, we can see that the generic partially-celestial scalar solution is parametrized by all possible $q^\mu$ that can be reached from $q^\mu_{{\rm ref}}$ by Lorentz transformations. These are one boost and $p-1$ rotations with angles $\alpha_k$ that take $q^\mu$ to a generic value $(\gamma,\gamma\vec{\beta})$. Hence the canonical Lorentz transformations are
\begin{eqnarray}\label{eq:gencel}
&&q^{\mu}=[L_q]^\mu_{~\nu}\, q^{\nu}_{{\rm ref}}\nonumber\\[5pt]
&&L_q=\prod_{k=1}^{D-3} R_{k,k+1}(\alpha_{k})\times B_{p}(\beta)\,.
\end{eqnarray}
Thus the partially-celestial scalar solution is parametrized by 1 boost, $D-2$ angles, and $D$ translations, for a total of $2D-1$ parameters of the coset $\mathcal{P}^D/LG^D$. One can easily check that this is indeed the correct dimension of this coset.
There is a one-to-one correspondence between the original $\phi^{\pm,\,PS}_{\Delta}$ and $\phi^{\pm, \,\mathcal{O}}_\Delta(x)$ for $\mathcal{O}\in\mathcal{P}^D/LG^D$. To see this, note that every $q^\mu$ in \eqref{eq:gencel} uniquely defines a $\vec{w}$, and vice versa.

\section{The Partially-Celestial Scalar Solution}
In this paper, we are interested in representations of the $D$-dimensional Poincar\'e group $\mathcal{P}^D=\mathbb{R}^D\rtimes SO(D-1,1)$ which transform covariantly under a $p+1$-dimensional Lorentz subgroup $SO(p,1)\subset \mathcal{P}^D$. These correspond to ordinary quantum fields on $p$-dimensional hypersurfaces in $d$ dimensions, which we call $p$-sheets for short. As a first step, we would like to construct the most general $SO(p,1)$ covariant solutions to \eqref{eq:weq}, drawing inspiration from the celestial solutions \eqref{eq:solsod1}. These would be akin to celestial solutions in $p$ ``sheet-parallel" dimensions, while being translationally invariant in $D-p$ ``external" directions. 
To this end, we define $p$\textit{-partially-celestial} scalars as
\begin{eqnarray}\label{eq:psscal}
&&\phi^{\pm}_{\Delta;p}(x;q,A)=\frac{2^{\frac{p-1}{2}+1}\pi^{\frac{p-1}{2}}}{(im)^{\frac{p-1}{2}}}\,\frac{(is)^{\alpha}}{\left[-q\cdot x\mp i\epsilon)\right]^\Delta}K_{\alpha}(ms)\nonumber\\[5pt]
&&s=\sqrt{x_\mu A^{\mu\nu} x_\nu},~~~\alpha=\Delta-\frac{p-1}{2}\,,
\end{eqnarray}
where $q^\mu$ is a null $D$-dimensional vector and $A^{\mu\nu}$ is a $D$-dimensional 2-index tensor. The latter is required to specify the embedding of a $p+1$-worldsheet in $D$-dimensional space. Similarly to our little group construction of fully celestial scalars in the previous section, here we need to specify reference values for both $A^{\mu\nu}_{p,{\rm ref}}$ and $q^\mu_{p,{\rm ref}}$. We choose
\begin{eqnarray}\label{eq:phiref}
&&\phi^{\pm,{\rm ref}}_{\Delta;p}(x)\equiv\phi^{\pm}_{\Delta;p}(x;q_{p,{\rm ref}},A_{p,{\rm ref}})\,,
\end{eqnarray}
where
\begin{eqnarray}\label{eq:projA}
&&A^{\mu\nu}_{p,{\rm ref}}\equiv \text{diag}\left(-1,1,\ldots,1,0,\ldots,0\right)\nonumber\\[5pt]
&&q^\mu_{p,{\rm ref}}\equiv \left(1,0,\ldots,0,1,0\ldots,0\right)\,.
\end{eqnarray}
where the $1$'s are repeated $p$-times in $A^{\mu\nu}_{p,{\rm ref}}$, while in $q^\mu_{p,{\rm ref}}$ the $1$ is in the $p+1$ entry. $A^{\mu\nu}_{p,{\rm ref}}$ is the projection tensor into the worldvolume of a $p$-hyperplane at rest, lying along the first $p$ dimensions, while $q^\mu_{p,{\rm ref}}$ is chosen so that $\phi^{\pm,{\rm ref}}_{\Delta;p}(x)$ is $SO(p-1)$ rotationally invariant, but it is also manifestly invariant under $M_{ip}-M_{0i},\,i\in\{1,\ldots,p-1\}$. It is also manifestly invariant under $\mathbb{R}^{D-p-1}\times SO(D-p-1)$ corresponding to ``external" translations and rotations. Overall, the little group under which $\phi^{\pm,{\rm ref}}_\Delta(x)$ is invariant is given by
\begin{eqnarray}\label{eq:LG}
LG^D_p=\mathbb{R}^{D-p-1}\times ISO(p-1)\times SO(D-p-1)\,,
\end{eqnarray}
where $dim(LG^D_p)=D-p-1+\frac{p(p-1)}{2}+\frac{(D-p-1)(D-p-2)}{2}$. For future reference, we also define for every $\Lambda\in SO(d-1,1)$ the Lorentz transformed tensor and vector,
\begin{eqnarray}\label{eq:qAt}
&&q^\mu_{p,\Lambda}\equiv \Lambda^\mu_{~\nu}q^\nu_{p,ref}\nonumber\\[5pt]
&&A^{\mu\nu}_{p,\Lambda}\equiv\Lambda^\mu_{~\alpha}\Lambda^\nu_{~\beta} A^{\alpha\beta}_{p,ref}\,.
\end{eqnarray}
Under Poincar\'e transformations, the partially-celestial scalar transforms as 
\begin{eqnarray}\label{eq:psscal}
&&\Omega=(\Lambda,v)\in\mathcal{P}^D:\nonumber\\[5pt]
&&\phi^{\pm, {\rm ref}}_{\Delta;p}(x)\rightarrow \phi^{\pm, \,\Omega}_{\Delta;p}(x)=\phi^{\pm, {\rm ref}}_{\Delta;p}(\Lambda(x+v))\nonumber\\[5pt]
&&=\phi^{\pm}_{\Delta;p}(x+v;q_{\Lambda^{-1} },A_\Lambda)\,.
\end{eqnarray}
Without loss of generality, we do the translation first in the Poincar\'e transformation. Importantly, not all Poincar\'e transformations actually lead to inequivalent solutions $\phi$ --- to label inequivalent solutions we have to mod out by the action of the little group. This is the same as finding canonical transformations $O\in \mathcal{P}^D/LG^D_p$. 
From \eqref{eq:psscal}, we can see that the generic partially-celestial scalar solution is parametrized by all possible $(q^\mu,A^{\mu\nu})$ that can be reached from $(q^\mu_{p,{\rm ref}},A^{\mu\nu}_{p,{\rm ref}})$ by Lorentz transformations.

To find the most generic $(q^\mu,A^{\mu\nu})$ we start from their reference values and perform a fixed set of independent Lorentz transformations. We start with transformations that act on $q^\mu$ within the $p+1$ dimensional reference hyperplane along time and the first p spatial directions, while leaving $A^{\mu\nu} invariant$. Since $q^\mu$ is a massless vector (even though our irreps are massive), we can get to a generic $q^\mu$ with 1 boost with velocity $\beta$ and $p-1$ rotations with angles $\alpha_k$. From now on, we can perform further transformations that act on $A^{\mu\nu}$, with $q^\mu$ going along for the ride. First, we can boost $A^{\mu\nu}$ by $\beta'$ to give is velocity (nonzero first row/column) in the $x_D$ direction. Without changing this velocity, we can perform rotations between all spatial directions orthogonal to $x_D$. However, out of these rotations, the ones outside the $p$-hyperplane leave the configuration invariant and are in fact part of the little group - that leaves $p(D-p-2)$ rotation angles $\varphi_{ij}$. At this point both $q^\mu$ and $A^{\mu\nu}$ are aligned in arbitrary directions orthogonal to $x_D$ and the velocity is in the $x_D$ direction. Finally, we can rotate the entire configuration arbitrarily in $D-1$ spatial directions, with $D-2$ angles $\theta_i$. In other words, the most general values for $(q^\mu,A^{\mu\nu})$ are $(q^\mu_{p,L_AL_q},A^{\mu\nu}_{p,L_A})$ where
\begin{eqnarray}\label{eq:genA}
&&L_q=\prod_{k=1}^{p-2} R_{k,k+1}(\alpha_{k})\times B_{p}(\beta)\,.\nonumber\\
&&L_A=\prod_{k=1}^{D-2} R_{k,k+1}(\theta_{k})\times\prod_{i=1}^{p}\prod_{j=p+1}^{D-2} R_{i,j}(\varphi_{ij})\times B_{D}(\beta')\,.\nonumber\\
\end{eqnarray}
Hence the partially-celestial scalar solution is parametrized by $(D-p-1)(p+1)-3$ angles, 2 boosts, and $p+1$ translations, for a total of $p(D-p)+D$ parameters of the coset $\mathcal{P}^D/LG^D_p$. One can easily check that this is indeed the correct dimension of this coset. To summarize, $p$-partially-celestial scalar solutions are solutions of the $D$-dimensional KG equation that are invariant under the little group (\ref{eq:LG}) and are labeled by $(D-p)(p+1)$ parameters of the coset $\mathcal{P}^D/LG^D_p$.

Finally, we note that the massless limit of the $p$-partially-celestial scalars \eqref{eq:phiref} is the same as that of a fully-celestial scalar. In other words, the information about a privileged $p$-hypersurface is lost in the massless limit.

\section{Partially-Celestial Scalars: Explicit Examples}
\subsection{Fully Celestial State in $D$ Dimensions}
This is the case discussed in Section~\ref{sec:PWCel}. One can easily check that setting $p=D-1$ in \eqref{eq:phiref}-\eqref{eq:genA}, we have $A^{\mu\nu}=\eta^{\mu\nu}$ which is Lorentz invariant, and so the partially celestial solution coincides with the definitions in Section~\ref{sec:PWCel}.

\subsection{Partially Celestial Line in $D$ Dimensions}
A partially celestial line in $D$ dimensions corresponds to a partially-celestial solution \eqref{eq:psscal} with $p=1$. Its little group is
\begin{eqnarray}
LG^D_1=R^2\times SO(D-2)\,,
\end{eqnarray}
whose dimension is $2+(D-2)(D-3)/2$. For $D=4$ we get $R^2\times SO(2)$. The coset $\mathcal{P}^D/LG^D_1$ has dimension $2D-1$, and it is parameterized by $D-3$ angles $\varphi_{ij}$, $D-2$ angles $\theta_k$, 2 boost parameters $\beta,\beta'$, one spatial translation and one time translation. The most generic $\phi^{\rm line}$ is given by
\begin{eqnarray}\label{eq:Tmunstr}
&&\phi^{\rm line}(x)=\phi^{\pm}_{\Delta;1}(x+a;q,A)\\[5pt]
&&a_{\nu}=(a_0,a_1,0,\ldots,0)\nonumber\\[5pt]
&&q_{\nu}=\sqrt{\frac{1+\beta}{1-\beta}}\,L_A\,q_{p,\rm ref}\nonumber\\[5pt]
&&A^{\mu\nu}=A^{\mu\nu}_{1,L_A}\nonumber\\[5pt]
&&L_A=\prod_{k=1}^{D-2} R_{k,k+1}(\theta_{k})\times\prod_{j=2}^{D-2} R_{1,j}(\varphi_{ij})\times B_{D}(\beta')\,.\nonumber
\end{eqnarray}

\subsection{Massive Particle in $D$ Dimensions}
A massive particle in $D$ dimensions corresponds to a partially-celestial solution \eqref{eq:psscal} with $p=0$. Since $q^\mu$ is ill-defined for $p=0$, the particle solution necessitates taking $\Delta=0$ rather than $\Delta=\frac{D-2}{2}+i\mathbb{R}$. In this case the little group (in Poincar\'e!) is
\begin{eqnarray}
LG^D_0=R^{D-1}\times SO(D-1)\,,
\end{eqnarray}
whose dimension is $D-1+(D-1)(D-2)/2$. For $D=4$ we get spatial tarnslations $R^3$ times the usual $SO(3)\simeq SU(2)$ little group for massive particles in 4D. The coset $\mathcal{P}^D/LG^D_0$ has dimension $D-1$, and it is parametrized by $D-2$ angles $\theta_k$, one boost parameter $\beta'$, and one time translation. The reference wavefunction $\phi^{particle}_{{\rm ref}}$ is given by
\begin{eqnarray}\label{eq:particleref}
&&\phi^{\rm particle}_{{\rm ref}}(x)=\phi^{\pm}_{\Delta;0}(x;q_{0,{\rm ref}},A_{0,{\rm ref}})=\frac{2i}{\pi}\sin(mt)\,.\nonumber\\
\end{eqnarray}
For particles, it is useful to define a slightly different reference state by
\begin{eqnarray}\label{eq:particleref}
\phi^{particle}_{\rm ref'}(x)&=&\frac{\pi}{2}\left[\phi^{particle}_{{\rm ref}}(x)-i\phi^{particle}_{{\rm ref}}(x+\delta x)\right]\nonumber\\[5pt]
&=&e^{imt}\,,
\end{eqnarray}
with $\delta x^\mu=\frac{\pi}{2m}(1,0,\ldots,0)$. This is simply a plane wave in the rest frame of the particle. In any other frame, we have
\begin{eqnarray}\label{eq:Tmunupart}
&&\phi^{particle}(x)=e^{im\gamma'(t+dt+\beta' \vec{x}\cdot\hat{n})}\nonumber\\[5pt]
&&\hat{n}=\prod_{k=1}^{D-2} R_{k,k+1}(\theta_{k})\,(0,\ldots,0,1)^T\,.
\end{eqnarray}

\section{From Partially-Celestial Scalars to Partially-Celestial Quantum States}
The $SO(p,1)$ invariance of partially-celestial scalars is suggestive of a new class of quantum states representing a $p$\textit{-sheet}. To make this correspondence more concrete, we can interpret the $p$-partially-celestial solution $\phi(x)$ as the wavefunction of a single $p$-partially-celestial state. 
First, we set up some notation. We denote canonical Poincar\'e transformations by $(L,a)\equiv O\in\mathcal{P}^D$ so that $[O]\in\mathcal{P}^D/LG^D_p$. As shown above, each canonical Poincar\'e transformation is labeled by $(D-p)(p+1)$ parameters. For every canonical transformation there is a unique partially-celestial scalar solution $\phi_O(x)=\phi_{{\rm ref}}(L(x+a))$. We can now identify
\begin{eqnarray}
\phi_O(x)=\left\langle 0\right|\Phi(x)a^\dagger_{O}\left|0\right\rangle\,,
\end{eqnarray}
where $\left|0\right\rangle$ is the vacuum, $a^\dagger_{O}$ is the creation operator for a $p$-partially-celestial scalar which is related to the reference scalar by the canonical Poincar\'e transforamtion $O$. $\Phi(x)$ is the usual field operator for the scalar field, which we can expand as
\begin{eqnarray}
\Phi(x)\equiv\int_{\mathcal{P}^D/LG^D_p} d O\,\left[\phi_O(x)a^\dagger_{O}+h.c.\right]\,.
\end{eqnarray}
As for the field operator for particles, we require the the field operator transforms covariantly under the Poincar\'e group \cite{Weinberg:1995mt}, 
\begin{eqnarray}\label{eq:Fieldtrans}
&&\Omega=(\Lambda,v^\mu)\in\mathcal{P}^D:\nonumber\\[5pt]
&&\Phi(x)\rightarrow U[\Omega]\Phi(x)U^{-1}[\Omega]=\Phi(\Lambda(x+v))\,.
\end{eqnarray}
In particular, $\Phi(x)$ is invariant under little group transformations $\Omega\in LG^D_p$. Similarly to particles, this requirement fixes the Poincar\'e transformation properties of the creation and annihilation operators (see derivation in Appendix~\ref{app:transf}),
\begin{eqnarray}\label{eq:optrans}
a^\dagger_{O}\rightarrow U[\Omega]a^\dagger_{O}U^{-1}[\Omega]=a^\dagger_{\Omega O}\,,
\end{eqnarray}
where the product $\Omega O$ is just the group product of the Poincar\'e group $\mathcal{P}^D$. In other words, a $p$-partially-celestial \textit{scalar state} $\left|O\right\rangle\equiv a^\dagger_O\left|0\right\rangle$ transforms as
\begin{eqnarray}
U[\Omega]\left|O\right\rangle=\left|\Omega O\right\rangle\,.
\end{eqnarray}
By construction, the reference state $\left|{\rm ref}\right\rangle=\left|O={\rm identity}\right\rangle$ is invariant under $\Omega\in LG^D_p$. This concludes our definition of the quantum state of a single $p$-partially-celestial scalar. 

\section{Partially-Celestial States with Spin: Wigner's Method}

In the previous section we defined $p$-partially-celestial scalar states by starting from a $p$-partially-celestial scalar solution and interpreting it as the wavefunction of a quantum state. Here we generalize our construction to $p$-partially-celestial states with spin. A direct generalization of our previous derivation would have been to start with a partially-celestial solution with spin, $\phi(x)_{i_1,\ldots i_l}$ and interpret it as the wavefunction for a spinning $p$-partially-celestial state (see for example, the constructions of fully-celestial spinors in \cite{Pasterski2017a,Iacobacci2020,Narayanan2020}) and massless $p$-forms in \cite{Donnay2022}. Instead, we will follow a simpler route, using Wigner's method of induced representations. Similarly to the scalar case, spinning states are labeled by $\left|O;\sigma\right\rangle$ where $O\in\mathcal{P}^D/LG^D_p$ and $\sigma$ is a composite spin index. The reference state is defined as usual as $\left|{\rm ref};\sigma\right\rangle=\left|O={\rm identity};\sigma\right\rangle$, and it is annihilated by all of the generators of the little group $LG^D_p$.  
Clearly, $LG^D_p$ is generated by the Poincar\'e algebra generators $G_n=\left\{M_{0i},M_{ip},P_k,M_{kl}\right\}$ where $i\in[1,\ldots p-1]$ and $k,l\in[p+1,\ldots d]$, so that $G_n \left|{\rm ref};\sigma\right\rangle = 0$. Consequently, the reference $p$-partially-celestial state $\left|{\rm ref}\right\rangle$ transforms in a representation of $LG^D_p$, i.e.
\begin{eqnarray}
U\left[W\right]\left|{\rm ref};\sigma\right\rangle=\mathcal{D}_{\sigma'\sigma}\left[W\right]\left|{\rm ref};\sigma'\right\rangle\,,
\end{eqnarray}
for any $W\in LG^D_p$. Here $\mathcal{D}_{\sigma'\sigma}\left[W\right]$ is some representation matrix of $LG^D_p$ and $\sigma$ is a collective index denoting a ``spin" label for $LG^D_p$ representations. For particles, $p=0$, and $D=4$, the representation matrices reduce to the normal spin representation matrices. 

As in Wigner's method for particles, for any $O\in\mathcal{P}^D/LG^D_p$ we can define the quantum state in a general frame as
\begin{eqnarray}\label{eq:Ostate}
\left|O;\sigma\right\rangle\equiv U\left[O\right]\left|{\rm ref};\sigma\right\rangle\,.
\end{eqnarray}
This also serves as a definition of $U\left[O\right]$ that can be uniquely extended to all $U\left[\Omega\right],\,\Omega\in\mathcal{P}^D$ acting on generic states. But first, let us ask ourselves which generators annihilate $\left|O\right\rangle$. The answer is straightforward. Take $O=(L,a)\in\mathcal{P}^D/LG^D_p$ and define $M^O_{\mu\nu}$ and $P^O_{\mu}$ so that
\begin{eqnarray}
M_{\mu\nu}&=&L^{~\alpha}_\mu L^{~\beta}_\nu\,\left(M^O_{\alpha\beta}-a_{[\alpha}P^O_{\beta]}\right)\nonumber\\[5pt]
P_{\mu}&=&L^{~\nu}_{\mu}P^O_\nu\,.
\end{eqnarray}
From chapter 2 of Weinberg's QFT book \cite{Weinberg:1995mt}, we have
\begin{eqnarray}
M_{\mu\nu}&=&U^{-1}\left[O\right]M^O_{\mu\nu}U\left[O\right]\nonumber\\[5pt]
P_{\mu}&=&U^{-1}\left[O\right]P^O_{\mu}U\left[O\right]\,.
\end{eqnarray}
Then $G^O_n\left|O;\sigma\right\rangle=0$.

Next, we can ask how the state $\left|O;\sigma\right\rangle$ transforms under a generic Poincar\'e transformation $\Omega\in\mathcal{P}^D$. This is uniquely defined using Wigner's method. Explicitly,
\begin{eqnarray}
U\left[\Omega\right]\left|O;\sigma\right\rangle&=&U\left[\bar{O}\right]U\left[\bar{O}^{-1}\Omega O\right]\left|{\rm ref};\sigma\right\rangle\nonumber\\[5pt]
&=&U\left[\bar{O}\right]U\left[W\right]\left|{\rm ref};\sigma\right\rangle\nonumber\\[5pt]
&=&\mathcal{D}_{\sigma'\sigma}\left[W\right]\left|\bar{O};\sigma'\right\rangle\,,
\end{eqnarray}
where $\bar{O}\in \mathcal{P}^D/LG^D_p$ is unique canonical Lorentz transformation defined by
\begin{eqnarray}
&&\bar{O}^\mu_{~\alpha}\,\bar{O}^\nu_{~\beta}\,A^{\alpha\beta}_{p,ref}=A^{\mu\nu}_{p,\Omega O}\nonumber\\[5pt]
&&\bar{O}^\mu_{~\alpha}q^{\alpha}_{p,ref}=q^{\mu}_{p,\Omega O}\,.
\end{eqnarray}
Note that generically $\Omega O$ is not a canonical Lorentz transformation in and of itself, and so $\bar{O}\neq \Omega O$. To conclude, in this section, we have straightforwardly applied Wigner's method of induced representations to single $p$-partially-celestial states with spin. The spin here is given by the representation $\mathcal{D}_{\sigma'\sigma}$ of the little group $LG^D_p$.

\section{Spinor Helicity Variables for $p$-sheet Scattering}\label{sec:spinors}

 Once we have fixed the little group for $p$-sheets we can construct the generalizations of the spinor-helicity variables, which are the key for the construction of the scattering amplitudes. 
Consider the compact part of the little group \eqref{eq:LG}:
\begin{eqnarray}\label{eq:LGc}
cLG^D_p=SO(p-1)\times SO(D-p-1)\,.
\end{eqnarray}
Our task is to define $D$-dimensional massive spinor-helicity variables that transform under $cLG^D_p$. We define two kinds of Minkowski spinors under the full $D$-dimensional Lorentz group, which also carry spinor indices under the (Euclidean) $SO$ factors of the  $cLG^D_p$. The first $\ket{L_A}_\alpha$ transforms with the little group $SO(D-p-1)$ spinor index, while the second $\kett{L_a}_\alpha$ transforms with an $SO(p-1)$ spinor index. For even $D$ the spinor representation is chiral, and we also have spinors of the opposite chirality $\sket{L^A}^{\dot{\alpha}}$ and $\skett{L^a}^{\dot{\alpha}}$. For even $D-p-1$ or $p-1$, we also have spinors with dotted little group indices. Undotted spinor indices are contracted in the northwest-southeast convention, while dotted ones are in the southwest-northeast convention.
We begin with a definition of $\ket{L_A}_\alpha,\,\sbra{L_{\dot{A}}}_{\dot{\alpha}}$, where $A,\,\dot{A}$ are $SO(D-p-1)$ spinor indices and $\alpha,\,\dot{\alpha}$ are $SO(D-1,1)$ spinor indices. Defining for any $N$, $s_N=2^{\lfloor N/2\rfloor-1} $, we have $A,\dot{A}=\{1,\ldots,s_{D-p-1}\}$ and $\alpha,\dot{\alpha}=\{1,\ldots,s_{D}\}$. 
The reference values for the single angle spinors are
\begin{eqnarray}\label{eq:lambdaref}
&&\ket{\rm ref_A}_\alpha=\delta_{A+s,\alpha}~~,~~\ket{\rm ref_{\dot{A}}}_\alpha=\delta_{\dot{A}+s,\alpha}\nonumber\\[5pt]
&&\sbra{\rm ref_{\dot{A}}}_{\dot{\alpha}}=\delta_{\dot{A}+s,\dot{\alpha}}~~,~~\sbra{\rm ref_{A}}_{\dot{\alpha}}=\delta_{A+s,\dot{\alpha}}\,,
\end{eqnarray}
where $s=s_{D}-s_{D-p-1}$. Note that the dotted Lorentz indices exist only for even dimensional $D$ and the dotted little group indices only exist for even dimensional $D-p-1$. For odd $D$, we can now define Lorentzian $D$-dimensional $[\Gamma^\mu]^{~\beta}_{\alpha}$ matrices, while for even $D$ we have the corresponding $[\Sigma^\mu]_{\alpha\dot{\beta}},\,[\bar{\Sigma}^\mu]^{\dot{\alpha}\beta}$ matrices. Similarly, for odd $D-p-1$ we have $D-p-1$ dimensional Euclidean $[\gamma_I]_{A}^{~B}$ matrices, while for even $D-p-1$ we have the corresponding $[\sigma^I]_{A\dot{B}},\,[\bar{\sigma}^I]^{\dot{A}B}$ matrices. We can always choose a basis so that the bottom right $s_{D-p-1}\times s_{D-p-1}$ block of the last $D-p-1$ $\Gamma/\Sigma$ matrices is numerically identical to the $\gamma/\sigma$ matrices. 
We can freely raise and lower the indices on the spinors via
\begin{eqnarray}\label{eq:rl}
&&\bra{{\rm ref}^A}^{\alpha}=\varepsilon^{\alpha\beta}\,\epsilon^{AB}\,\ket{{\rm ref}_B}_{\beta}\nonumber\\[5pt]
&&\sket{{\rm ref}^{\dot{A}}}^{\dot{\alpha}}=\sbra{{\rm ref}_{\dot{B}}}_{\dot{\beta}}\,\epsilon^{\dot{B}\dot{A}}\,\varepsilon^{\dot{\beta}\dot{\alpha}}\,,
\end{eqnarray}
where $\varepsilon=i\,[\Gamma^{D-2}]$ for odd $D$ and $\varepsilon=i\,[\Sigma^{D-2}]$ for even $D$ and similarly for $\epsilon$ with $\gamma$ and $\sigma$.

Now, we can use these gamma/sigma matrices to combine the spinors into 
\begin{eqnarray}\label{eq:Alambdaref}
&&A^{\mu\nu}_{p,\rm ref}=\eta^{\mu\nu}\,-\,[v_{\rm ref}]^{\{\mu}_I~ [v_{\rm ref}]^{\nu\};I}\nonumber\\[10pt]
&&\text{odd}~D,~~\text{odd}~D-p-1:\nonumber\\[5pt]
&&[v_{\rm ref}]^{\mu}_I=\bra{\rm ref^A}^{\alpha}~[\Gamma^\mu]^{~\beta}_{\alpha}~[\gamma_I]_{A}^{~B}~\ket{\rm ref_B}_{\beta}\nonumber\\[10pt]
&&\text{even}~D,~~\text{even}~D-p-1:\nonumber\\[5pt]
&&[v_{\rm ref}]^{\mu}_I=\bra{\rm ref^A}^{\alpha}~[\Sigma^\mu]_{\alpha\dot{\beta}}~[\sigma_I]_{A\dot{B}}~\sket{\rm ref^{\dot{B}}}^{\dot{\beta}}\,,
\end{eqnarray}
and similarly for mixed parity $D$ and $D-p-1$. In analogy with Wigner's method, we define $\ket{L_A}_{\alpha},\,\sbra{L_{\dot{A}}}_{\dot{\alpha}},\,$ in any other frame as
\begin{eqnarray}\label{eq:lambdaoth}
&&\ket{L_A}_{\alpha}=L_\alpha^{~\beta}\,\ket{\rm ref_A}_{\beta}\,\nonumber\\[5pt]
&&\sbra{L_{\dot{A}}}_{\dot{\alpha}}=\sbra{\rm ref_{\dot{A}}}_{\dot{\beta}}L_{~\dot{\alpha}}^{\dot{\beta}}\,\,
\end{eqnarray}
for every $L\in SO(D-1,1)/cLG^D_p$. One can readily check that
\begin{eqnarray}\label{eq:Alambda}
&&A^{\mu\nu}_{p,L}=\eta^{\mu\nu}\,-\,[v_{L}]^{\{\mu}_I~ [v_L]^{\nu\};I}\nonumber\\[10pt]
&&\text{odd}~D,~~\text{odd}~D-p-1:\nonumber\\[5pt]
&&[v_L]^{\mu}_I=\bra{L^A}^{\alpha}~[\Gamma^\mu]^{~\beta}_{\alpha}~[\gamma_I]_{A}^{~B}~\ket{L_B}_{\beta}\nonumber\\[10pt]
&&\text{even}~D,~~\text{even}~D-p-1:\nonumber\\[5pt]
&&[v_{L}]^{\mu}_I=\bra{L^A}^{\alpha}~[\Sigma^\mu]_{\alpha\dot{\beta}}~[\sigma_I]_{A\dot{B}}~\sket{L^{\dot{B}}}^{\dot{\beta}}\,,
\end{eqnarray}
In fact the first equation can also be thought of as the definition of the spinor-helicity variables. It is the generalization of the relation $p=|p\rangle [p|$ for the definition of the ordinary spinors.

By Wigner's method, $\ket{L_A}_{\alpha},\,\sbra{L_{\dot{A}}}_{\dot{\alpha}}$ transform under a generic $\Lambda\in SO(D-1,1)$ as
\begin{eqnarray}\label{eq:lambdatr}
&&\Lambda_\alpha^{~\beta}\,\ket{L_A}_{\beta}=\bar{L}_\alpha^\beta\, W_\beta^{~\gamma}\,\ket{\rm ref_A}_{\gamma}\,\nonumber\\[5pt]
&&\sbra{L_{\dot{A}}}_{\dot{\beta}}\,~\Lambda^{\dot{\beta}}_{~\dot{\alpha}}\,=\,\sbra{L_{\dot{A}}}_{\dot{\gamma}}\,W^{\dot{\gamma}}_{~\dot{\beta}}\,\bar{L}^{\dot{\beta}}_{~\dot{\alpha}} \,,
\end{eqnarray}
where $\bar{L}\in SO(D-1,1)/cLG^D_p$ is the unique canonical Lorentz transformation defined by $A_{p,\Lambda L}=\bar{L}A_{p,\rm ref}\bar{L}^T$, and $W=\bar{L}^{-1}\Lambda L \in SO(D-p-1)$ is a little group transformation. Now, by the definition \eqref{eq:lambdaref}, we have  
\begin{eqnarray}\label{eq:lambdatr}
&&W_\beta^{~\gamma}\,\ket{\rm ref_A}_{\gamma}=W_{A}^{~B}\,\ket{{\rm ref}_B}_{\beta}\,\nonumber\\[5pt]
&&\sbra{\rm ref_{\dot{A}}}_{\dot{\gamma}}W^{\dot{\gamma}}_{~\dot{\beta}}\,=\sbra{{\rm ref}_{\dot{B}}}_{\dot{\beta}}\,W^{\dot{B}}_{~\dot{A}}\,.
\end{eqnarray}
In other words, when acting on the reference spinors with a spacetime-index little group transformation is the same as acting on them with a same transformation in the little group indices. This is the same thing that happens to massive spinors in 4D \cite{Arkani-Hamed:2017jhn}. We conclude that
\begin{eqnarray}\label{eq:lambdatr2}
&&\Lambda_\alpha^{~\beta}\,\ket{L_A}_{\beta}=W_{A}^{~B}\,\ket{\bar{L}_B}_{\alpha}\,\nonumber\\[5pt]
&&\sbra{L_{\dot{A}}}_{\dot{\beta}}\,\Lambda^{\dot{\beta}}_{~\dot{\alpha}}=\sbra{\bar{L}_{\dot{B}}}_{\dot{\alpha}}\,\,W^{\dot{B}}_{~\dot{A}}\,,
\end{eqnarray}
i.e. these spinors transform exactly with the correct $SO(D-p-1)$ little group factor. That makes them the right building blocks for $p$-partially-celestial amplitudes. 

Similarly, we can define the spinors $\kett{{\rm ref}^a}_{\alpha},\,\sbraa{{\rm ref}_{a}}_{\dot{\alpha}}$ where $a$ is an $SO(p-1)$ little group index. Note that we do not dot the a index for reasons that will become apparent momentarily. These are defined as
\begin{eqnarray}\label{eq:etaref}
&&\kett{\rm ref_a}_\alpha=\delta^{a\alpha}\nonumber\\[5pt]
&&\sbraa{\rm ref^{a}}_{\dot{\alpha}}=\delta^a_{\dot{\alpha}}\,.
\end{eqnarray}
They are defined so that
\begin{eqnarray}\label{eq:qetarefodd}
&&q^{\mu}_{p,\rm ref}=\braa{\rm ref^{a}}^{\alpha}~[\Gamma^\mu]^{~\beta}_{\alpha}~\kett{\rm ref_{a}}_{\beta}\,.
\end{eqnarray}
for odd $D$ and $p-1$, and 
\begin{eqnarray}\label{eq:qetarefeven}
&&q^{\mu}_{p,\rm ref}=\braa{\rm ref^{a}}^{\alpha}~[\Sigma^\mu]_{\alpha\dot{\beta}}~\skett{\rm ref_{a}}^{\dot{\beta}}\,,
\end{eqnarray}
for even $D$ and $p-1$.
Similarly to $\ket{L},\,\sbra{L}$, the generic $\kett{L},\,\sbraa{L}$ transform as
\begin{eqnarray}\label{eq:etatr2}
&&\Lambda_\alpha^{~\beta}\,\kett{L_a}_{\beta}=W_{a}^{~b}\,\kett{\bar{L}'_b}_{\alpha}\nonumber\\[5pt]
&&\sbraa{L^a}_{\dot{\beta}}\Lambda^{\dot{\beta}}_{~\dot{\alpha}}=\sbraa{\bar{L}'^b}_{\dot{\alpha}}\,W_b^{~a}\,,
\end{eqnarray}
where $\bar{L}'\in SO(D-1,1)/cLG^D_p$ is the unique canonical Lorentz transformation defined by $q_{p,\Lambda L}=\bar{L}'q_{p,\rm ref}$, and $W=\bar{L}^{'\,-1}\Lambda L\in SO(p-1)$ is a little group transformation. One can readily check that
\begin{eqnarray}\label{eq:qetaodd}
&&q^{\mu}_{p,L}=\braa{L^{a}}^{\alpha}~[\Gamma^\mu]^{~\beta}_{\alpha}~\kett{L_{a}}_{\beta}\,.
\end{eqnarray}
for odd $D$ and $p-1$, and 
\begin{eqnarray}\label{eq:qetaeven}
&&q^{\mu}_{p,L}=\braa{L^{a}}^{\alpha}~[\Sigma^\mu]_{\alpha\dot{\beta}}~\skett{L_{a}}^{\dot{\beta}}\,,
\end{eqnarray}
for even $D$ and $p-1$, for any $L\in SO(D-1,1)/cLG^D_p$. Again these last two relations can be thouhgt of as the definitions of double-line $SO(p-1)$ spinors.
We then see that $\kett{L},\,\sbraa{L}$ transform with the correct $SO(p-1)$ little group transformation and can be used to form $p$-partially-celestial amplitudes.   

\section{Constructing Partially Celestial Amplitudes}\label{sec:3pt}
Using the spinor-helicity variables defined in the previous section, we can construct the little group-covariant part of any partially-celestial amplitude, generalizing the 4D massive formalism of \cite{Arkani-Hamed:2017jhn}.
We take all external states to all live in $D$-dimensional space and have ``internal dimensions" $p_n$, with $n=1,2,3,\ldots,N$ and representations $\mathcal{R}^{in}_n\times \mathcal{R}^{out}_n$ under the compact little group $cLG^D_{p_n}=SO(p_n-1)\times SO(D-p_n-1)$. To saturate the required little group transformation of the amplitude, we need to combine the little group indices of the spinor-helicity variables $\ket{n_{A_n}}_{\alpha_n},\,\sbra{n_{\dot{A}_n}}_{\dot{\alpha}_n},\,\kett{n_{a_n}}_{\alpha_n}$, and $\sbraa{n_{a_n}}_{\dot{\alpha}_n}$. This is achieved via contractions of the $cLG^D_{p_n}$ indices using the $\gamma/\sigma$ matrices. Once the correct little group transformation is obtained, all Lorentz indices can be contracted via the little group invariants $\varepsilon^{\alpha\beta},\,\varepsilon_{\dot{\alpha}\dot{\beta}}$, $[nk]_{~\alpha}^{\beta}\equiv [n^I]_{\alpha \dot{\gamma}}[k_I]^{\dot{\gamma}\beta}$ and $[nk]_{\dot{\alpha}}^{~\dot{\beta}}\equiv[n^I]_{\dot{\alpha}\gamma }[k_I]^{\gamma\dot{\beta}}$ where
\begin{eqnarray}\label{eq:nI}
[n^I]_{\alpha \dot{\beta}}=\ket{n_A}_\alpha \epsilon^{AB}\sigma^I_{B\dot{C}}\epsilon^{\dot{C}\dot{D}}\sbra{n_{\dot{D}}}_{\dot{\beta}}\,,
\end{eqnarray}
as well as their double angle / double square bracket counterparts.

As an example, consider the (little group covariant part of) the 3-pt amplitude for three 3-celestial-amplitudes in 10D, transforming as the $(0,\mathbf{4}),\,(0,\mathbf{\bar{4}})$ and $(0,\mathbf{6})$ of $cLG^{10}_3=U(1)\times SO(6)$. This amplitude is given by
\begin{eqnarray}\label{eq:amp3b}
\mathcal{A}^{A_1,\dot{A}_2,I_3}=\bks{3_{\dot{A}_3}2^{\dot{A}_2} }[\bar{\sigma}^{I_3}]^{\dot{A}_3B_3}\bk{1^{A_1} 3_{B_3}}\,,
\end{eqnarray}
where all the spinors are defined for $D=10$ and $p=3$. Another example is the (little group covariant part of) the 3-pt amplitude for a line (1-partially-celestial) state emitting a massive scalar particle in 4D. We consider the case in which the two 1-partially-celestial legs have helicity $\pm\frac{1}{2}$ under $cLG^{4}_1=SO(2)\simeq U(1)$ (in the all-incoming convention). The amplitude in this case is 
\begin{eqnarray}\label{eq:amp2st1p}
\mathcal{A}^{A_1,A_2}=\bk{1^{A_1} 2^{A_2}}\,.
\end{eqnarray}
where both the spinors are defined for $D=4$ and $p=1$, and the values of $A_1,\,A_2$ correspond to different choices of positive or negative helicity. We can get a direct analytical expression for this amplitude using the explicit values of the spinors given in Appendix~\ref{app:explicit}. As an illustration, consider a line at rest along the x-axis emitting a massive scalar particle, while remaining at rest and rotating by an angle $\varphi$. The amplitude for this process is
\begin{eqnarray}\label{eq:amp2st1p}
\mathcal{A}^{A_1,A_2}=\colmatt{0&e^{-\frac{i\varphi}{2}}\\-e^{\frac{i\varphi}{2}}&0}\,.
\end{eqnarray}
Note that the amplitude is helicity conserving in the all-incoming convention.

Finally, note that translational invariance should pose additional constraints on partially-celestial amplitudes. In fact, for particles we know that translational invariance (i.e. momentum conservation) dictates that 3-pt amplitudes are completely fixed by their little group transformations. In our case, similarly to the case of fully celestial amplitudes \cite{Stieberger2019,Law2020a}, the generators for translation assume a nonlinear differential form when expressed in terms of $(q,A,\Delta)$. We leave the exploration of the constraints of translational invariance on partially celestial amplitudes for future work, including whether or not 3-pt partially-celestial amplitudes are fixed by their little group transformations.

\section{Pairwise Little Group}
Consider a scalar $p$-sheet parallel to a scalar $p'$ sheet in $D$ dimensions. To be parallel, we require that $p+p'\leq D-2$. By applying Poincar\'e transformations, we can always go to the ``center of velocity" frame of the two sheets, in which their wavefunctions are given by
\begin{eqnarray}
&&\phi^{\pm,{\rm ref}}_{\Delta;p}(x)=\phi^{\pm}_{\Delta;p}(x;q_{\rm ref,p},B_{\rm ref,p})\nonumber\\[5pt]
&&\phi^{\pm,{\rm ref}}_{\Delta';p'}(x)=\phi^{\pm}_{\Delta';p'}(x;q_{\rm ref,p'},B_{\rm ref,p'})\nonumber\\[5pt]
&&B^{\mu\nu}_{\rm ref,p}\,=\,\text{diag}(0,1,\ldots,1,0,\ldots,0,0,\ldots,0)+M_\perp(\beta)\nonumber\\[5pt]
&&B^{\mu\nu}_{\rm ref,p'}=\text{diag}(0,0,\ldots,0,1,\ldots,1,0,\ldots,0)+M_\perp(-\beta)\nonumber\\[5pt]
&&M_\perp(\beta)=-\left(\begin{array}{ccccc}\gamma^2&0&\ldots&0&\gamma^2\beta\\0&0&0&0&0\\\vdots&0&\ddots&0&\vdots\\0&0&0&0&0\\\ \gamma^2\beta&0&\ldots&0&\gamma^2\beta^2\end{array}\right)
\,.
\end{eqnarray}
In analogy with the pairwise little group for particles, we can ask which subgroup of $\mathcal{P}^D$ stabilizes both $\phi^{\pm,{\rm ref}}_{\Delta;p}(x)$ and $\phi^{\pm,{\rm ref}}_{\Delta';p'}(x)$. The answer is
\begin{eqnarray}\label{eq:PairwiseLG}
pLG^D_{p,p'}=&&ISO(p-1)\times ISO(p'-1)\times\nonumber\\[5pt]
&&R^{D-p-p'-2}\times SO(D-p-p'-2)\,.\nonumber\\
\end{eqnarray}
For particles in 4D, $p=p'=0$, and the pairwise little group reduces to $SO(2)\simeq U(1)$, consistently with \cite{Csaki:2020yei,Csaki:2020inw,Csaki2022,Csaki2022a} (see also \cite{Lippstreu2021} for a discussion of pairwise helicity in the context of 4D celestial amplitudes). In particular, we focus on the case in which $p'=D-p-4$. This is the case where the sheets are \textit{mutually-non-local}, in the sense that they source $p$-form gauge fields that are EM-dual to each other. In this case the $p$-form charges of the two sheets are constrained by Dirac quantization, as shown in \cite{Polchinski:1995mt},
\begin{eqnarray}\label{eq:DiracQ}
q\equiv eg=\frac{n}{2},~~~p\neq D-p-4\,,
\end{eqnarray}
where $e$ and $g$ are the charge of the $p$ and $D-p-4$ sheets, respectively.
In the self-dual case, $p=\frac{D-4}{2}$, the sheets can be dyonic, and the Dirac quantization condition is generalized to \cite{Henneaux:1986ht}
\begin{eqnarray}\label{eq:DiracQSD}
q\equiv e_1g_2+(-1)^pe_2g_1=\frac{n}{2},~~~p=\frac{D-4}{2}\,.
\end{eqnarray}
In \cite{Csaki:2020yei,Csaki:2020inw,Csaki2022,Csaki2022a}, the Dirac-quantized quantities $q$ were shown to play the role of pairwise helicities labeling the representations of the $U(1)$ little group. Here the situation is similar; substituting $p'=D-p-4$ in \eqref{eq:PairwiseLG}, we have
\begin{eqnarray}\label{eq:PairwiseLGDual}
&&pLG^D_{p,D-p-4}=ISO(p-1)\times ISO(D-p-5)\times\nonumber\\[5pt]
&&R^{2}\times U(1)\,.
\end{eqnarray}
and we see that indeed $pLG^D_{p,D-p-4}$ has a $U(1)$ factor. We can naturally identify the pairwise helicities labeling the representations of this factor of the pairwise little group with the $q$ given in \eqref{eq:DiracQ}-\eqref{eq:DiracQSD}.

The presence of a $U(1)$ factor for the pairwise little group for mutually-non-local partially-celestial states hints that the entire structure exposed in \cite{Csaki:2020yei,Csaki:2020inw,Csaki2022,Csaki2022a} generalizes directly to the present case. This is reminiscent of a pair of mutually-non-local branes, which source  $p$-form  and $p'$-form fields and thus carry extra angular momentum in these fields. This extra angular momentum modifies the selection rules for brane scattering, in the same way it modifies them for monopoles and charges in 4D. 

\section{Outlook and Future Work}

In this paper we defined the quantum states for scalar and spinning $p$-partially-celestial states  and, notably, the generalized $LG^D_p$-covariant spinor-helicity variables in $D$ dimensions. These results allow us to find the little group covariant part of the most general 3-point amplitudes for partially-celestial states. Additionally we found the corresponding pairwise little group, which has a $U(1)$ factor for mutually-non-local  states. In a future little group construction for branes, the helicities under the pairwise little group should be identified with Dirac quantized products of charges  by examining the Lorentz-transformation properties of soft-photon-dressed electric and magnetic states as in ref. \cite{Csaki2022}; We expect that the same result can be shown for branes by considering their ``soft-higher-gauge field" dressed multi-brane states. The generalized spinor-helicity variables enable the bottom-up construction of (the little group covariant part of) scattering amplitudes for $p$-partially-celestial states. We give a procedure for constructing the little group covariant part of 3-pt functions for 3-partially-celestial states in 10D and for two lines and a scalar particle in 4D. Unlike in the case of particles, we cannot be sure whether or not 3-pt amplitudes for partially celestial states are completely fixed by their little group transformation. We leave that question for future work, in which we will analyze in detail the constraints from translational invariance on partially-celestial amplitudes. 
%%%%%%%%%%%%%%%%%%%%%%%%%%%%%%%%%%%%%%%%%%%%%%%%%%%%%%
\section*{Acknowledgements}
We thank Yale Fan for initial collaboration on ideas related to this work. While the resulting papers turned out quite different, there is some overlap in the aims and methods of this paper and~\cite{Fan:2022svn}. CC is supported in part by the NSF grant PHY-2014071 as well as the US-Israeli BSF grant 2016153.  OT was supported in part by the DOE under grant DE-AC02-05CH11231.
J.T. is supported by the DOE under grant  DE-SC-0009999. 
\appendix
\section{From Field Operator to Ladder Operators}\label{app:transf}
Here we show that the transformation properties of the field operator \eqref{eq:Fieldtrans}, lead to the transformation \eqref{eq:optrans} of the creation operator. To see this, write for $\Omega=(\Lambda,v)\in\mathcal{P}^D$,
\begin{eqnarray}\label{eq:Fieldtrans2}
&&\Phi(\Lambda(x+v))=\nonumber\\[5pt]
&&\int_{\mathcal{P}^D/LG^D_p} d O\,\left[\phi_O(x)\,U[\Omega]a^\dagger_{O}U^{-1}[\Omega]+h.c.\right]\,,
\end{eqnarray}
or in other words
\begin{eqnarray}\label{eq:Fieldtrans3}
&&\int_{\mathcal{P}^D/LG^D_p} d O'\,\left[\phi_{O'}(\Lambda x+v)\,a^\dagger_{O'}+h.c.\right]=\nonumber\\[5pt]
&&\int_{\mathcal{P}^D/LG^D_p} d O\,\left[\phi_O(x)\,U[\Omega]a^\dagger_{O}U^{-1}[\Omega]+h.c.\right]\,.
\end{eqnarray}
Note that $\phi_{O'}(\Lambda x+v)=\phi_{\Omega^{-1}O'}(x)$, and we can change the integration variable on the LH side as $O'=PO$, we have
\begin{eqnarray}\label{eq:Fieldtrans3}
&&\int_{\mathcal{P}^D/LG^D_p} d O\,\left[\phi_{O}(x)a^\dagger_{\Omega O}+h.c.\right]=\nonumber\\[5pt]
&&\int_{\mathcal{P}^D/LG^D_p} d O\,\left[\phi_O(x)U[\Omega]a^\dagger_{O}U^{-1}[\Omega]+h.c.\right]\,,
\end{eqnarray}
from which \eqref{eq:optrans} follows.

\section{Explicit Parametrizations}\label{app:explicit}
For completeness, we present here the generic $A^{\mu\nu},\lambda,\widetilde{\lambda}$ for a massive particle in 4D, and the generic $q^\mu,\,A^{\mu\nu},\lambda,\widetilde{\lambda}$ for a line-partially-celestial state in 4D. 
\subsection{Massive Particle}
As a special case of \eqref{eq:Tmunstr}, the generic wavefunction for a massive particle in 4D depends on the translation $(a_0,0,0,0)$, one boosts $\beta'$ and two angles $\theta_{1},\,\theta_{2}$. The most general $A^{\mu\nu}$ is then
\begin{eqnarray}\label{eq:Apart4d}
&& A^{\mu\nu}=- u^{\mu} u^{\nu}\,.
\end{eqnarray}
where $u^\mu$ is defined the same way as \eqref{eq:us}, and is the 4-velocity of the particle, given by
\begin{eqnarray}\label{eq:us}
&&u^{\mu}=\gamma'(-1,\beta'\sin\theta_{2}\sin\theta_1,\beta'\sin\theta_{2}\cos\theta_1,\beta'\cos\theta_{1})\,.\nonumber\\
\end{eqnarray}
Finally, we define the spinors $\ket{L},\,\bra{L}$ corresponding to a massive particle in 4D. By the definitions in section~\ref{sec:spinors}, they are given by
\begin{eqnarray}
&&\ket{L^A}_{\alpha}={\left(\sbra{L^{A}}_{\dot{\alpha}}\right)}^{*}=\,\nonumber\\[5pt]
&&\colmatt{e^{\frac{i\theta_{1}}{2}}a'_-\cos\left(\frac{\theta_{2}}{2}\right)&ie^{\frac{i\theta_{1}}{2}}a'_+\sin\left(\frac{\theta_{2}}{2}\right)\\ie^{-\frac{i\theta_{1}}{2}}a'_-\sin\left(\frac{\theta_{2}}{2}\right)&e^{-\frac{i\theta_{1}}{2}}a'_+\cos\left(\frac{\theta_{2}}{2}\right)}\,,
\end{eqnarray}
where $a'_\pm=\sqrt{\gamma'(1\pm\beta')}$. Note that we do not have dotted little group indices since the little group is $SO(3)\simeq SU(2)$ whose $\mathbf{2}$ and $\bar{\mathbf{2}}$ are equivalent. One can readily check that \eqref{eq:Alambda} is satisfied.

\subsection{Line}
As a special case of \eqref{eq:Tmunstr}, the generic wavefunction for a line-partially-celestial state in 4D depends on the translations $(a_0,a_1,0,0)$, two boosts $\beta,\beta'$ and three angles $\theta_{1},\,\theta_{2},\,\varphi_{12}$. The most general $(A^{\mu\nu},q^\mu)$ are then
\begin{eqnarray}\label{eq:qstring4d}
&& A^{\mu\nu}=\xi^{\mu}\xi^{\nu} - u^{\mu} u^{\nu} \nonumber\\[5pt]
&& q^{\mu}=\sqrt{\frac{1+\beta}{1-\beta}}\,\left(u^{\mu}+\xi^{\mu}\right)\,.
\end{eqnarray}
where $u^\mu$ is the same 4-velocity given in \eqref{eq:us}, while $\xi^\mu$ is given by
\begin{eqnarray}\label{eq:dus}
&&\xi^{\mu}=(0,\cos \theta_1 \cos \varphi _{12}-\sin
   \theta_{1} \cos \theta_{2} \sin \varphi_{12},\nonumber\\[5pt]
   &&~~~~~~~~-\sin \theta_{1}\cos \varphi_{12}-\cos
   \theta_{1} \cos \theta_{2} \sin \varphi_{12},\nonumber\\[5pt]
   &&~~~~~~~~~\sin \theta_{2} \sin \varphi_{12}\})\,.
\end{eqnarray}
Note that $\xi^{\mu}$ denotes the line's 4-orientation, which is always transverse to the 4-velocity, $u\cdot \xi=0$.
Finally, we define the spinors $\ket{L},\,\bra{L}$ corresponding to the generic line in 4D. By the definitions in section~\ref{sec:spinors}, they are given by
\begin{widetext}
\begin{eqnarray}
&&\ket{L^A}_{\alpha}={\left(\sbra{L^{\dot{A}}}_{\dot{\alpha}}\right)}^{*}=\nonumber\\[5pt]
&&\left(
\begin{array}{cc}
 e^{\frac{i \theta_1}{2}} \left[a_+ a'_- e^{\frac{i \varphi_{12}}{2}} \cos \left(\frac{\theta_2}{2}\right)-i a_- a'_+ e^{-\frac{i \varphi_{12}}{2}} \sin \left(\frac{\theta_2}{2}\right)\right] & e^{\frac{i \theta_1}{2}} \left[i a_+ a'_+ e^{-\frac{i \varphi_{12}}{2}} \sin \left(\frac{\theta_2}{2}\right)-a_- a'_- e^{\frac{i \varphi_{12}}{2}} \cos \left(\frac{\theta_2}{2}\right)\right] \\ e^{-\frac{i \theta_1}{2}} \left[i a_+ a'_- e^{\frac{i \varphi_{12}}{2}} \sin \left(\frac{\theta_2}{2}\right)-a_- a'_+ e^{-\frac{i \varphi_{12}}{2}} \cos \left(\frac{\theta_2}{2}\right)\right] & e^{-\frac{i \theta_1}{2}} \left[a_+ a'_+ e^{-\frac{i \varphi_{12}}{2}} \cos \left(\frac{\theta_2}{2}\right)-i a_- a'_- e^{\frac{i \varphi_{12}}{2}} \sin \left(\frac{\theta_2}{2}\right)\right] \\
\end{array}
\right)\,,
\end{eqnarray}
\end{widetext}
where $a_{\pm}=\sqrt{\frac{\gamma\pm 1}{2}}$. Note that we do not have dotted little group indices; the little group is $SO(2)\simeq U(1)$, and so the 2-component spinor representation is  \textit{reducible} and includes both $\pm\frac{1}{2}$ helicities under the $U(1)$. One can readily check that \eqref{eq:Alambda} is satisfied.

\bibliographystyle{apsrev4-1}
\bibliography{Refs}{}

\begin{thebibliography}{26}%
\makeatletter
\providecommand \@ifxundefined [1]{%
 \@ifx{#1\undefined}
}%
\providecommand \@ifnum [1]{%
 \ifnum #1\expandafter \@firstoftwo
 \else \expandafter \@secondoftwo
 \fi
}%
\providecommand \@ifx [1]{%
 \ifx #1\expandafter \@firstoftwo
 \else \expandafter \@secondoftwo
 \fi
}%
\providecommand \natexlab [1]{#1}%
\providecommand \enquote  [1]{``#1''}%
\providecommand \bibnamefont  [1]{#1}%
\providecommand \bibfnamefont [1]{#1}%
\providecommand \citenamefont [1]{#1}%
\providecommand \href@noop [0]{\@secondoftwo}%
\providecommand \href [0]{\begingroup \@sanitize@url \@href}%
\providecommand \@href[1]{\@@startlink{#1}\@@href}%
\providecommand \@@href[1]{\endgroup#1\@@endlink}%
\providecommand \@sanitize@url [0]{\catcode `\\12\catcode `\$12\catcode
  `\&12\catcode `\#12\catcode `\^12\catcode `\_12\catcode `\%12\relax}%
\providecommand \@@startlink[1]{}%
\providecommand \@@endlink[0]{}%
\providecommand \url  [0]{\begingroup\@sanitize@url \@url }%
\providecommand \@url [1]{\endgroup\@href {#1}{\urlprefix }}%
\providecommand \urlprefix  [0]{URL }%
\providecommand \Eprint [0]{\href }%
\providecommand \doibase [0]{http://dx.doi.org/}%
\providecommand \selectlanguage [0]{\@gobble}%
\providecommand \bibinfo  [0]{\@secondoftwo}%
\providecommand \bibfield  [0]{\@secondoftwo}%
\providecommand \translation [1]{[#1]}%
\providecommand \BibitemOpen [0]{}%
\providecommand \bibitemStop [0]{}%
\providecommand \bibitemNoStop [0]{.\EOS\space}%
\providecommand \EOS [0]{\spacefactor3000\relax}%
\providecommand \BibitemShut  [1]{\csname bibitem#1\endcsname}%
\let\auto@bib@innerbib\@empty
%</preamble>
\bibitem [{\citenamefont {Pasterski}\ and\ \citenamefont
  {Shao}(2017)}]{Pasterski2017a}%
  \BibitemOpen
  \bibfield  {author} {\bibinfo {author} {\bibfnamefont {S.}~\bibnamefont
  {Pasterski}}\ and\ \bibinfo {author} {\bibfnamefont {S.-H.}\ \bibnamefont
  {Shao}},\ }\href {\doibase 10.1103/PhysRevD.96.065022} {\bibfield  {journal}
  {\bibinfo  {journal} {Phys. Rev. D}\ }\textbf {\bibinfo {volume} {96}},\
  \bibinfo {pages} {065022} (\bibinfo {year} {2017})},\ \Eprint
  {http://arxiv.org/abs/1705.01027} {arXiv:1705.01027 [hep-th]} \BibitemShut
  {NoStop}%
\bibitem [{\citenamefont {Pasterski}\ \emph
  {et~al.}(2017{\natexlab{a}})\citenamefont {Pasterski}, \citenamefont {Shao},\
  and\ \citenamefont {Strominger}}]{Pasterski2017b}%
  \BibitemOpen
  \bibfield  {author} {\bibinfo {author} {\bibfnamefont {S.}~\bibnamefont
  {Pasterski}}, \bibinfo {author} {\bibfnamefont {S.-H.}\ \bibnamefont {Shao}},
  \ and\ \bibinfo {author} {\bibfnamefont {A.}~\bibnamefont {Strominger}},\
  }\href {\doibase 10.1103/PhysRevD.96.065026} {\bibfield  {journal} {\bibinfo
  {journal} {Phys. Rev. D}\ }\textbf {\bibinfo {volume} {96}},\ \bibinfo
  {pages} {065026} (\bibinfo {year} {2017}{\natexlab{a}})},\ \Eprint
  {http://arxiv.org/abs/1701.00049} {arXiv:1701.00049 [hep-th]} \BibitemShut
  {NoStop}%
\bibitem [{()}]{}%
  \BibitemOpen
  \href@noop {} {\ }\BibitemShut {NoStop}%
\bibitem [{\citenamefont {Pasterski}\ \emph
  {et~al.}(2017{\natexlab{b}})\citenamefont {Pasterski}, \citenamefont {Shao},\
  and\ \citenamefont {Strominger}}]{Pasterski2017}%
  \BibitemOpen
  \bibfield  {author} {\bibinfo {author} {\bibfnamefont {S.}~\bibnamefont
  {Pasterski}}, \bibinfo {author} {\bibfnamefont {S.-H.}\ \bibnamefont {Shao}},
  \ and\ \bibinfo {author} {\bibfnamefont {A.}~\bibnamefont {Strominger}},\
  }\href {\doibase 10.1103/PhysRevD.96.085006} {\bibfield  {journal} {\bibinfo
  {journal} {Phys. Rev. D}\ }\textbf {\bibinfo {volume} {96}},\ \bibinfo
  {pages} {085006} (\bibinfo {year} {2017}{\natexlab{b}})},\ \Eprint
  {http://arxiv.org/abs/1706.03917} {arXiv:1706.03917 [hep-th]} \BibitemShut
  {NoStop}%
\bibitem [{\citenamefont {Kapec}\ and\ \citenamefont
  {Mitra}(2018)}]{Kapec2018}%
  \BibitemOpen
  \bibfield  {author} {\bibinfo {author} {\bibfnamefont {D.}~\bibnamefont
  {Kapec}}\ and\ \bibinfo {author} {\bibfnamefont {P.}~\bibnamefont {Mitra}},\
  }\href {\doibase 10.1007/JHEP05(2018)186} {\bibfield  {journal} {\bibinfo
  {journal} {JHEP}\ }\textbf {\bibinfo {volume} {05}},\ \bibinfo {pages} {186}
  (\bibinfo {year} {2018})},\ \Eprint {http://arxiv.org/abs/1711.04371}
  {arXiv:1711.04371 [hep-th]} \BibitemShut {NoStop}%
\bibitem [{\citenamefont {Banerjee}\ and\ \citenamefont
  {Pandey}(2020)}]{Banerjee2020}%
  \BibitemOpen
  \bibfield  {author} {\bibinfo {author} {\bibfnamefont {S.}~\bibnamefont
  {Banerjee}}\ and\ \bibinfo {author} {\bibfnamefont {P.}~\bibnamefont
  {Pandey}},\ }\href {\doibase 10.1007/JHEP02(2020)067} {\bibfield  {journal}
  {\bibinfo  {journal} {JHEP}\ }\textbf {\bibinfo {volume} {02}},\ \bibinfo
  {pages} {067} (\bibinfo {year} {2020})},\ \Eprint
  {http://arxiv.org/abs/1906.01650} {arXiv:1906.01650 [hep-th]} \BibitemShut
  {NoStop}%
\bibitem [{\citenamefont {Banerjee}\ \emph {et~al.}(2020)\citenamefont
  {Banerjee}, \citenamefont {Pandey},\ and\ \citenamefont
  {Paul}}]{Banerjee2020a}%
  \BibitemOpen
  \bibfield  {author} {\bibinfo {author} {\bibfnamefont {S.}~\bibnamefont
  {Banerjee}}, \bibinfo {author} {\bibfnamefont {P.}~\bibnamefont {Pandey}}, \
  and\ \bibinfo {author} {\bibfnamefont {P.}~\bibnamefont {Paul}},\ }\href
  {\doibase 10.1103/PhysRevD.101.106014} {\bibfield  {journal} {\bibinfo
  {journal} {Phys. Rev. D}\ }\textbf {\bibinfo {volume} {101}},\ \bibinfo
  {pages} {106014} (\bibinfo {year} {2020})},\ \Eprint
  {http://arxiv.org/abs/1902.02309} {arXiv:1902.02309 [hep-th]} \BibitemShut
  {NoStop}%
\bibitem [{\citenamefont {Pasterski}\ \emph {et~al.}(2021)\citenamefont
  {Pasterski}, \citenamefont {Pate},\ and\ \citenamefont
  {Raclariu}}]{Pasterski2021}%
  \BibitemOpen
  \bibfield  {author} {\bibinfo {author} {\bibfnamefont {S.}~\bibnamefont
  {Pasterski}}, \bibinfo {author} {\bibfnamefont {M.}~\bibnamefont {Pate}}, \
  and\ \bibinfo {author} {\bibfnamefont {A.-M.}\ \bibnamefont {Raclariu}},\
  }in\ \href@noop {} {\emph {\bibinfo {booktitle} {{2022 Snowmass Summer
  Study}}}}\ (\bibinfo {year} {2021})\ \Eprint
  {http://arxiv.org/abs/2111.11392} {arXiv:2111.11392 [hep-th]} \BibitemShut
  {NoStop}%
\bibitem [{\citenamefont {Cs\'aki}\ \emph
  {et~al.}(2021{\natexlab{a}})\citenamefont {Cs\'aki}, \citenamefont {Hong},
  \citenamefont {Shirman}, \citenamefont {Telem}, \citenamefont {Terning},\
  and\ \citenamefont {Waterbury}}]{Csaki:2020inw}%
  \BibitemOpen
  \bibfield  {author} {\bibinfo {author} {\bibfnamefont {C.}~\bibnamefont
  {Cs\'aki}}, \bibinfo {author} {\bibfnamefont {S.}~\bibnamefont {Hong}},
  \bibinfo {author} {\bibfnamefont {Y.}~\bibnamefont {Shirman}}, \bibinfo
  {author} {\bibfnamefont {O.}~\bibnamefont {Telem}}, \bibinfo {author}
  {\bibfnamefont {J.}~\bibnamefont {Terning}}, \ and\ \bibinfo {author}
  {\bibfnamefont {M.}~\bibnamefont {Waterbury}},\ }\href {\doibase
  10.1007/JHEP08(2021)029} {\bibfield  {journal} {\bibinfo  {journal} {JHEP}\
  }\textbf {\bibinfo {volume} {08}},\ \bibinfo {pages} {029} (\bibinfo {year}
  {2021}{\natexlab{a}})},\ \Eprint {http://arxiv.org/abs/2009.14213}
  {arXiv:2009.14213 [hep-th]} \BibitemShut {NoStop}%
\bibitem [{\citenamefont {Cs\'aki}\ \emph
  {et~al.}(2021{\natexlab{b}})\citenamefont {Cs\'aki}, \citenamefont {Hong},
  \citenamefont {Shirman}, \citenamefont {Telem},\ and\ \citenamefont
  {Terning}}]{Csaki:2020yei}%
  \BibitemOpen
  \bibfield  {author} {\bibinfo {author} {\bibfnamefont {C.}~\bibnamefont
  {Cs\'aki}}, \bibinfo {author} {\bibfnamefont {S.}~\bibnamefont {Hong}},
  \bibinfo {author} {\bibfnamefont {Y.}~\bibnamefont {Shirman}}, \bibinfo
  {author} {\bibfnamefont {O.}~\bibnamefont {Telem}}, \ and\ \bibinfo {author}
  {\bibfnamefont {J.}~\bibnamefont {Terning}},\ }\href {\doibase
  10.1103/PhysRevLett.127.041601} {\bibfield  {journal} {\bibinfo  {journal}
  {Phys. Rev. Lett.}\ }\textbf {\bibinfo {volume} {127}},\ \bibinfo {pages}
  {041601} (\bibinfo {year} {2021}{\natexlab{b}})},\ \Eprint
  {http://arxiv.org/abs/2010.13794} {arXiv:2010.13794 [hep-th]} \BibitemShut
  {NoStop}%
\bibitem [{\citenamefont {Cs\'aki}\ \emph
  {et~al.}(2022{\natexlab{a}})\citenamefont {Cs\'aki}, \citenamefont {Shirman},
  \citenamefont {Telem},\ and\ \citenamefont {Terning}}]{Csaki2022a}%
  \BibitemOpen
  \bibfield  {author} {\bibinfo {author} {\bibfnamefont {C.}~\bibnamefont
  {Cs\'aki}}, \bibinfo {author} {\bibfnamefont {Y.}~\bibnamefont {Shirman}},
  \bibinfo {author} {\bibfnamefont {O.}~\bibnamefont {Telem}}, \ and\ \bibinfo
  {author} {\bibfnamefont {J.}~\bibnamefont {Terning}},\ }\href {\doibase
  10.1103/PhysRevLett.129.181601} {\bibfield  {journal} {\bibinfo  {journal}
  {Phys. Rev. Lett.}\ }\textbf {\bibinfo {volume} {129}},\ \bibinfo {pages}
  {181601} (\bibinfo {year} {2022}{\natexlab{a}})}\BibitemShut {NoStop}%
\bibitem [{\citenamefont {Cs\'aki}\ \emph
  {et~al.}(2022{\natexlab{b}})\citenamefont {Cs\'aki}, \citenamefont {Dong},
  \citenamefont {Telem}, \citenamefont {Terning},\ and\ \citenamefont
  {Yankielowicz}}]{Csaki2022}%
  \BibitemOpen
  \bibfield  {author} {\bibinfo {author} {\bibfnamefont {C.}~\bibnamefont
  {Cs\'aki}}, \bibinfo {author} {\bibfnamefont {Z.-Y.}\ \bibnamefont {Dong}},
  \bibinfo {author} {\bibfnamefont {O.}~\bibnamefont {Telem}}, \bibinfo
  {author} {\bibfnamefont {J.}~\bibnamefont {Terning}}, \ and\ \bibinfo
  {author} {\bibfnamefont {S.}~\bibnamefont {Yankielowicz}},\ }\href@noop {} {\
   (\bibinfo {year} {2022}{\natexlab{b}})},\ \Eprint
  {http://arxiv.org/abs/2209.03369} {arXiv:2209.03369 [hep-th]} \BibitemShut
  {NoStop}%
\bibitem [{\citenamefont {Law}\ and\ \citenamefont
  {Zlotnikov}(2020{\natexlab{a}})}]{Law2020}%
  \BibitemOpen
  \bibfield  {author} {\bibinfo {author} {\bibfnamefont {Y.~T.~A.}\
  \bibnamefont {Law}}\ and\ \bibinfo {author} {\bibfnamefont {M.}~\bibnamefont
  {Zlotnikov}},\ }\href {\doibase 10.1007/JHEP06(2020)079} {\bibfield
  {journal} {\bibinfo  {journal} {JHEP}\ }\textbf {\bibinfo {volume} {06}},\
  \bibinfo {pages} {079} (\bibinfo {year} {2020}{\natexlab{a}})},\ \Eprint
  {http://arxiv.org/abs/2004.04309} {arXiv:2004.04309 [hep-th]} \BibitemShut
  {NoStop}%
\bibitem [{\citenamefont {Banerjee}(2019)}]{Banerjee2019}%
  \BibitemOpen
  \bibfield  {author} {\bibinfo {author} {\bibfnamefont {S.}~\bibnamefont
  {Banerjee}},\ }\href {\doibase 10.1007/JHEP01(2019)205} {\bibfield  {journal}
  {\bibinfo  {journal} {JHEP}\ }\textbf {\bibinfo {volume} {01}},\ \bibinfo
  {pages} {205} (\bibinfo {year} {2019})},\ \Eprint
  {http://arxiv.org/abs/1801.10171} {arXiv:1801.10171 [hep-th]} \BibitemShut
  {NoStop}%
\bibitem [{\citenamefont {Donnay}\ \emph {et~al.}(2022)\citenamefont {Donnay},
  \citenamefont {Esmaeili},\ and\ \citenamefont {Heissenberg}}]{Donnay2022}%
  \BibitemOpen
  \bibfield  {author} {\bibinfo {author} {\bibfnamefont {L.}~\bibnamefont
  {Donnay}}, \bibinfo {author} {\bibfnamefont {E.}~\bibnamefont {Esmaeili}}, \
  and\ \bibinfo {author} {\bibfnamefont {C.}~\bibnamefont {Heissenberg}},\
  }\href@noop {} {\  (\bibinfo {year} {2022})},\ \Eprint
  {http://arxiv.org/abs/2212.03060} {arXiv:2212.03060 [hep-th]} \BibitemShut
  {NoStop}%
\bibitem [{\citenamefont {Donnay}\ \emph {et~al.}(2019)\citenamefont {Donnay},
  \citenamefont {Puhm},\ and\ \citenamefont {Strominger}}]{Donnay2019}%
  \BibitemOpen
  \bibfield  {author} {\bibinfo {author} {\bibfnamefont {L.}~\bibnamefont
  {Donnay}}, \bibinfo {author} {\bibfnamefont {A.}~\bibnamefont {Puhm}}, \ and\
  \bibinfo {author} {\bibfnamefont {A.}~\bibnamefont {Strominger}},\ }\href
  {\doibase 10.1007/JHEP01(2019)184} {\bibfield  {journal} {\bibinfo  {journal}
  {JHEP}\ }\textbf {\bibinfo {volume} {01}},\ \bibinfo {pages} {184} (\bibinfo
  {year} {2019})},\ \Eprint {http://arxiv.org/abs/1810.05219} {arXiv:1810.05219
  [hep-th]} \BibitemShut {NoStop}%
\bibitem [{\citenamefont {Stieberger}\ and\ \citenamefont
  {Taylor}(2019)}]{Stieberger2019}%
  \BibitemOpen
  \bibfield  {author} {\bibinfo {author} {\bibfnamefont {S.}~\bibnamefont
  {Stieberger}}\ and\ \bibinfo {author} {\bibfnamefont {T.~R.}\ \bibnamefont
  {Taylor}},\ }\href {\doibase 10.1016/j.physletb.2019.03.063} {\bibfield
  {journal} {\bibinfo  {journal} {Phys. Lett. B}\ }\textbf {\bibinfo {volume}
  {793}},\ \bibinfo {pages} {141} (\bibinfo {year} {2019})},\ \Eprint
  {http://arxiv.org/abs/1812.01080} {arXiv:1812.01080 [hep-th]} \BibitemShut
  {NoStop}%
\bibitem [{\citenamefont {Law}\ and\ \citenamefont
  {Zlotnikov}(2020{\natexlab{b}})}]{Law2020a}%
  \BibitemOpen
  \bibfield  {author} {\bibinfo {author} {\bibfnamefont {Y.~T.~A.}\
  \bibnamefont {Law}}\ and\ \bibinfo {author} {\bibfnamefont {M.}~\bibnamefont
  {Zlotnikov}},\ }\href {\doibase 10.1007/JHEP03(2020)085} {\bibfield
  {journal} {\bibinfo  {journal} {JHEP}\ }\textbf {\bibinfo {volume} {03}},\
  \bibinfo {pages} {085} (\bibinfo {year} {2020}{\natexlab{b}})},\ \bibinfo
  {note} {[Erratum: JHEP 04, 202 (2020)]},\ \Eprint
  {http://arxiv.org/abs/1910.04356} {arXiv:1910.04356 [hep-th]} \BibitemShut
  {NoStop}%
\bibitem [{\citenamefont {Weinberg}(2005)}]{Weinberg:1995mt}%
  \BibitemOpen
  \bibfield  {author} {\bibinfo {author} {\bibfnamefont {S.}~\bibnamefont
  {Weinberg}},\ }\href@noop {} {\emph {\bibinfo {title} {{The Quantum theory of
  fields. Vol. 1: Foundations}}}}\ (\bibinfo  {publisher} {Cambridge University
  Press},\ \bibinfo {year} {2005})\BibitemShut {NoStop}%
%%CITATION = INSPIRE-406190;%%
\bibitem [{\citenamefont {Iacobacci}\ and\ \citenamefont
  {M\"uck}(2020)}]{Iacobacci2020}%
  \BibitemOpen
  \bibfield  {author} {\bibinfo {author} {\bibfnamefont {L.}~\bibnamefont
  {Iacobacci}}\ and\ \bibinfo {author} {\bibfnamefont {W.}~\bibnamefont
  {M\"uck}},\ }\href {\doibase 10.1103/PhysRevD.102.106025} {\bibfield
  {journal} {\bibinfo  {journal} {Phys. Rev. D}\ }\textbf {\bibinfo {volume}
  {102}},\ \bibinfo {pages} {106025} (\bibinfo {year} {2020})},\ \Eprint
  {http://arxiv.org/abs/2009.02938} {arXiv:2009.02938 [hep-th]} \BibitemShut
  {NoStop}%
\bibitem [{\citenamefont {Narayanan}(2020)}]{Narayanan2020}%
  \BibitemOpen
  \bibfield  {author} {\bibinfo {author} {\bibfnamefont {S.~A.}\ \bibnamefont
  {Narayanan}},\ }\href {\doibase 10.1007/JHEP12(2020)074} {\bibfield
  {journal} {\bibinfo  {journal} {JHEP}\ }\textbf {\bibinfo {volume} {12}},\
  \bibinfo {pages} {074} (\bibinfo {year} {2020})},\ \Eprint
  {http://arxiv.org/abs/2009.03883} {arXiv:2009.03883 [hep-th]} \BibitemShut
  {NoStop}%
\bibitem [{\citenamefont {Arkani-Hamed}\ \emph {et~al.}(2017)\citenamefont
  {Arkani-Hamed}, \citenamefont {Huang},\ and\ \citenamefont
  {Huang}}]{Arkani-Hamed:2017jhn}%
  \BibitemOpen
  \bibfield  {author} {\bibinfo {author} {\bibfnamefont {N.}~\bibnamefont
  {Arkani-Hamed}}, \bibinfo {author} {\bibfnamefont {T.-C.}\ \bibnamefont
  {Huang}}, \ and\ \bibinfo {author} {\bibfnamefont {Y.-t.}\ \bibnamefont
  {Huang}},\ }\href@noop {} {\  (\bibinfo {year} {2017})},\ \Eprint
  {http://arxiv.org/abs/1709.04891} {arXiv:1709.04891 [hep-th]} \BibitemShut
  {NoStop}%
\bibitem [{\citenamefont {Lippstreu}(2021)}]{Lippstreu2021}%
  \BibitemOpen
  \bibfield  {author} {\bibinfo {author} {\bibfnamefont {L.}~\bibnamefont
  {Lippstreu}},\ }\href {\doibase 10.1007/JHEP11(2021)051} {\bibfield
  {journal} {\bibinfo  {journal} {JHEP}\ }\textbf {\bibinfo {volume} {11}},\
  \bibinfo {pages} {051} (\bibinfo {year} {2021})},\ \Eprint
  {http://arxiv.org/abs/2106.00084} {arXiv:2106.00084 [hep-th]} \BibitemShut
  {NoStop}%
\bibitem [{\citenamefont {Polchinski}(1995)}]{Polchinski:1995mt}%
  \BibitemOpen
  \bibfield  {author} {\bibinfo {author} {\bibfnamefont {J.}~\bibnamefont
  {Polchinski}},\ }\href {\doibase 10.1103/PhysRevLett.75.4724} {\bibfield
  {journal} {\bibinfo  {journal} {Phys. Rev. Lett.}\ }\textbf {\bibinfo
  {volume} {75}},\ \bibinfo {pages} {4724} (\bibinfo {year} {1995})},\ \Eprint
  {http://arxiv.org/abs/hep-th/9510017} {arXiv:hep-th/9510017} \BibitemShut
  {NoStop}%
\bibitem [{\citenamefont {Henneaux}\ and\ \citenamefont
  {Teitelboim}(1986)}]{Henneaux:1986ht}%
  \BibitemOpen
  \bibfield  {author} {\bibinfo {author} {\bibfnamefont {M.}~\bibnamefont
  {Henneaux}}\ and\ \bibinfo {author} {\bibfnamefont {C.}~\bibnamefont
  {Teitelboim}},\ }\href {\doibase 10.1007/BF01889624} {\bibfield  {journal}
  {\bibinfo  {journal} {Found. Phys.}\ }\textbf {\bibinfo {volume} {16}},\
  \bibinfo {pages} {593} (\bibinfo {year} {1986})}\BibitemShut {NoStop}%
\bibitem [{\citenamefont {Fan}(2022)}]{Fan:2022svn}%
  \BibitemOpen
  \bibfield  {author} {\bibinfo {author} {\bibfnamefont {Y.}~\bibnamefont
  {Fan}},\ }\href@noop {} {\  (\bibinfo {year} {2022})},\ \Eprint
  {http://arxiv.org/abs/2212.05725} {arXiv:2212.05725 [hep-th]} \BibitemShut
  {NoStop}%
\end{thebibliography}%
\end{document}